\documentclass[twocolumn,amssymb,amsmath,floats,showpacs,superscriptaddress,pre]{revtex4-1}
\usepackage{amsmath,amssymb,bm}
\usepackage{graphics}
\usepackage{epsfig}
\usepackage{graphicx}

\begin{document}

\title{Belief-propagation algorithm and the Ising model on networks with arbitrary distributions of motifs}

\author{S. Yoon}
\affiliation{Departamento de F{\'\i}sica da Universidade de Aveiro,
I3N, 3810-193 Aveiro, Portugal}
\author{A. V. Goltsev}
\affiliation{Departamento de F{\'\i}sica da Universidade de Aveiro,
I3N, 3810-193 Aveiro, Portugal} \affiliation{A.~F. Ioffe
Physico-Technical Institute, 194021 St. Petersburg, Russia}
\author{S. N. Dorogovtsev}
\affiliation{Departamento de F{\'\i}sica da Universidade de Aveiro,
I3N, 3810-193 Aveiro, Portugal} \affiliation{A.~F. Ioffe
Physico-Technical Institute, 194021 St. Petersburg, Russia}
\author{J. F. F. Mendes}
\affiliation{Departamento de F{\'\i}sica da Universidade de Aveiro,
I3N, 3810-193 Aveiro, Portugal}

\date{\today}

\begin{abstract}
We generalize the belief-propagation algorithm to sparse random
networks with arbitrary distributions of motifs (triangles, loops,
etc.). Each vertex in these networks belongs to a given set of
motifs (generalization of the configuration model). These networks
can be treated as sparse uncorrelated hypergraphs in which
hyperedges represent motifs. Here a hypergraph is a generalization
of a graph, where a hyperedge can connect any number of vertices.
These uncorrelated hypergraphs are tree-like (hypertrees), which
crucially simplifies the problem and allows us to apply the
belief-propagation algorithm to these loopy networks with arbitrary
motifs. As natural examples, we consider motifs in the form of
finite loops and cliques. We apply the belief-propagation algorithm
to the ferromagnetic Ising model with pairwise interactions on the resulting random networks and
obtain an exact solution of this model.
%% on networks with finite loops or cliques as motifs.
We find an exact
critical temperature of the ferromagnetic phase transition and
demonstrate that with increasing the clustering coefficient and the
loop size, the critical temperature increases compared to ordinary
tree-like complex networks. However, weak clustering does not change the critical behavior qualitatively. Our solution also gives the birth point
of the giant connected component in these loopy networks.

\end{abstract}

\maketitle

%%%%%%%\email{sdorogov@fis.ua.pt}

%%%%%%%\email{goltsev@fis.ua.pt}

%%%%%%%\email{jfmendes@fis.ua.pt}

%%\date{}

\section{Introduction}
\label{intro}

The belief-propagation algorithm is an effective numerical method
for solving inference problems on sparse graphs. It was originally
proposed by J. Pearl \cite{Pearl:p88} for tree-like graphs. The
belief-propagation algorithm was applied to study diverse systems in
computer science, physics, and biology. Among its numerous
applications are computer vision problems, decoding of high
performance turbo codes and many others, see
\cite{Frey:98,McEliece:mmc98}. Empirically, it was found that this
algorithm works surprisingly good even for graphs with loops.
J.~S.~Yedidia \emph{et al.} \cite{Yedidia:yfw01} found that the
belief-propagation algorithm actually coincides with the
minimization of the Bethe free energy. This discovery renewed
interest in the Bethe-Peierls approximation and related methods
\cite{Pretti:pp03,Mooij:mk05,Hartmann:hw05}. In statistical physics
the belief-propagation algorithm is equivalent to the so-called
\emph{cavity method} proposed by M\'ezard,  Parisi, and Virasoro
\cite{mp2003}.
%%The recent
%%progress in the survey propagation algorithm, which was introduced
%%to solve some difficult combinatorial optimization problems, is a
%%good example of interference between computer science and
%%statistical physics \cite{Mezard:mpz02,Mezard:mz02,Braunstein:bz04}.
The belief-propagation algorithm is applicable to models with both
discrete and continuous variables
\cite{mooij04,ohkubo05,Dorogovtsev:dgm08}. Recently the
belief-propagation algorithm was applied to diverse problems in
random networks: spread of disease  \cite{karrer10}, the
calculation of the size of the giant component \cite{sk10}, counting
large loops in directed random uncorrelated networks
\cite{bianconi08}, the graph bi-partitioning problem \cite{sulc10},
analysis of the contributions of individual nodes and groups of
nodes to the network structure \cite{ras10}, and networks of spiking
neurons \cite{smd09}.

These investigations showed that the belief-propagation algorithm is
exact for a model on a graph with locally tree-like structure in the
infinite size limit. However, only an approximate solution was
obtained by use of this method for networks with short loops. Real
technological, social and biological networks have numerous short and large loops and other complex
subgraphs or motifs which lead to essentially non-tree-like
neighborhoods \cite{ab01a,dm01c,Dorogovtsev:2010,Newman:n03a,Milo2002,Milo2004,Sporn2004,Alon2007}.
That is why it is important to develop a method which
takes into account finite loops and more complex motifs. Different
ways were proposed recently to compute loop corrections to the Bethe
approximation by use of the belief-propagation algorithm
\cite{mr05,ps05}. These methods, however, are exact only if a graph
contains a single loop. Loop expansions were proposed in Refs.
\cite{Chertkov:cc06a, Chertkov:cc06b}. However, they were not applied to complex
network yet.

Recently, Newman and Miller \cite{newman:n09,m09,kn10b}
independently introduced a model of random graphs with arbitrary
distributions of subgraphs or motifs.
%%, for example, triangles, finite loops, etc.
In this natural generalization of the configuration model, each
vertex belongs to a given set of motifs [e.g., vertex $i$ is a
member of $Q^{(1)}(i)$ motifs of type $1$, $Q^{(2)}(i)$ motifs of
type $2$, and so on], and apart from this constraint, the network is
uniformly random. In the original configuration model, the sequence
of vertex degrees $Q(i)$ is fixed, where $i=1,2,...,N$. In this
generalization, the sequence of generalized vertex degrees
$Q^{(1)}(i),Q^{(2)}(i),...$ is given. For example, motifs can be
triangles, loops, chains, cliques (fully connected subgraphs),
single edges that do not enter other motifs, and, in general,
arbitrary finite clusters. The resulting networks can be treated as
uncorrelated hypergraphs in which hyperedges represent motifs. In
graph theory, a hypergraph is a generalization of a graph, where a
hyperedge can connect any number of vertices \cite{Berge73}. Because
of the complex motifs, the large sparse networks under consideration
can have loops, clustering, and correlations, but the underlying
hypergraphs are locally tree-like (hypertrees) and uncorrelated
similarly to the original sparse configuration model. Our approach
is based on the hypertree-like structure of these highly structured
sparse networks. To demonstrate our approach, we apply the
generalized belief-propagation algorithm to the ferromagnetic Ising
model on the sparse networks in which motifs are finite loops or
cliques. We obtain an exact solution of the ferromagnetic Ising
model and the birth point of the giant connected component in this
kind of highly structured networks. Note that the Ising model on
these networks is a more complex problem than the percolation
problem because we must account for spin interactions between spins
both inside and between motifs. We find an exact critical
temperature of the ferromagnetic phase transition and demonstrate
that finite loops increase the critical temperature in comparison to
ordinary tree-like networks.

\section{Ensemble of random networks with motifs}

%%Configuration model of uncorrelated random hypergraphs}
\label{config}

\begin{figure}[t]
\begin{center}
\scalebox{0.6}{\includegraphics[angle=0]{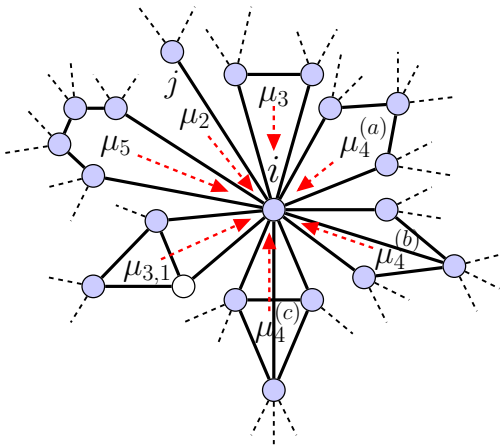}}
\end{center}
\caption{(Color online) Different motifs can be attached to vertex
$i$ in a hypergraph: an ordinary edge $a_{ji}$, a triangle, a
square, and a pentagon. There are also non-symmetric motifs consisting
of four vertices and a clique of size 4. A motif can have internal
vertices (open circle) that have edges only inside this motif.
%%two usual edges, $a_{ji}$ and $a_{j'i}$. There are
%%also two triangles, a square, and a hexagon which are attached to
%%this vertex.
Motifs form a locally tree-like hypergraph where they play a role of
hyperedges. Arrows represent incoming messages that arrive at
vertex $i$ from the motifs attached to this vertex. } \label{fig1}
\end{figure}

Let us introduce a statistical ensemble of random networks with a
given distribution of motifs in which each vertex belongs to a given
set of motifs and apart of this constraint, the networks are
uniformly random. These networks can be treated as uncorrelated
hypergraphs, in which motifs play a role of hyperedges, so the
number of hyperdegrees are equal to the number of specific motifs
attached to a vertex. In principal, one can choose any subgraph with
an arbitrary number of vertices as a motif, see Fig.~\ref{fig1}. In
the present paper, for simplicity, we only consider simple motifs
such as single edges, finite loops, and cliques.

Let us first note how one can describe a statistical ensemble of
random networks with the simplest motifs, namely simple edges (see,
for example, Refs. \cite{Dorogovtsev:dgm08, Bianconi09} and
references therein). We define the probability $p_{2}(a_{ij})$ that
an edge between vertices $i$ and $j$ is present ($a_{ij}=1$) or
absent ($a_{ij}=0$),
\begin{equation}
p_{2}(a_{ij})=\frac{\left\langle Q_{2} \right\rangle }{N-1}\delta
(a_{ij}-1)+\Bigl(1-\frac{\left\langle Q_{2} \right\rangle }{N-1}\Bigr)\delta (a_{ij})
\label{edge}
\end{equation}
where $a_{ij}$ are entries of the symmetrical adjacency matrix,
$\langle Q_{2} \rangle = \langle Q \rangle$ is the mean number of
edges attached to a vertex.
%%%%%%%%%%%%%%%%%%%%%%%%%%%%%%%%%%%%%%%%%%%%%%%%
It is well known that the probability of
the realization of the Erd\H{o}s-R\'{e}nyi graph with a given adjacency matrix
$a_{ij}$, is the product
\begin{equation}
G_{2}(\{a_{ij}\}) =\prod_{i=1}^{N-1} \prod_{j=i+1}^{N}
p_{2}(a_{ij}). \label{edge2}
\end{equation}
The degree distribution %%$P_{2}(Q_{2})$
is the Poisson distribution.
%%%%%%%%%%%%%%%%%%%%%%%%%%%%%%%%%%%%%%%%%%%%%%%%%%%%%
In the configuration model, the probability of the realization of a
given graph with a given sequence of degrees, $Q_{2}(1)$,
$Q_{2}(2)$, $Q_{2}(3)$,\dots, $Q_{2}(N)\equiv \{Q_{2}(i)\}$, is
\begin{equation}
G_{2}(\{a_{ij}\})=\frac{1}{A}\prod_{i=1}^{N}\delta \Bigl(Q_{2}(i) -
\sum_{j=1}^{N}a_{ij}\Bigr)\prod_{i<j}p_{2}(a_{ij}).
\label{edge-config-model}
\end{equation}
%%Here $P_{2}(a_{ij})$ is given by Eq.~(\ref{edge}).
The delta-function fixes the number of edges attached to vertex $i$.
$A$ is a normalization constant. The distribution function of
degrees is determined by the sequence $\{Q_{2}(i)\}$,
\begin{equation}
P_{2}(Q_{2})=\frac{1}{N}\sum_{i}\delta(Q_{2}-Q_{2}(i)).
\label{dd-edges}
\end{equation}

The second simplest motif, the triangle, plays the role of a hyperedge
that interconnects a triple of vertices. Let us introduce the
probability $p_{3}(a_{ijk})$ that a hyperedge (triangle) among
vertices $i$, $j$, and $k$ is present or absent, i.e., $a_{ijk}=1$
or $a_{ijk}=0$, respectively:
\begin{equation}
p_{3}(a_{ijk})=p \delta (a_{ijk}-1)+(1-p) \delta
(a_{ijk}),  \label{s-edge}
\end{equation}
where
\begin{equation}
p=\frac{2\left\langle Q_{3} \right\rangle}{(N-1)(N-2)} \label{p3}
\end{equation}
is the probability that vertices $i,j$ and $k$ form a triangle.
$\left\langle Q_{3} \right\rangle $ is the mean number of triangles
attached to a randomly chosen vertex. $a_{ijk}$ are entries of the
adjacency matrix of the hypergraph. This matrix is symmetrical with
respect to permutations of the indices $i,j$ and $k$,
$a_{ijk}=a_{jki}=a_{kij}=\dots$\,\,.

For example, one can introduce the ensemble of the Erd\H{o}s-R\'{e}nyi
hypergraphs. Given that the matrix elements
$a_{ijk}$ are independent and uncorrelated random parameters, the
probability of realization of a graph with a given adjacency matrix
$a_{ijk}$ is the product of probabilities $p_{3}(a_{ijk})$ over
different triples of vertices:
\begin{equation}
G_{3}(\{a_{ijk}\})=\prod_{i=1}^{N-2}\prod_{j=i+1}^{N-1}\prod_{k=j+1}^{N}
p_{3}(a_{ijk}). \label{s-ER-graph}
\end{equation}
This function describes the ensemble of the Erd\H{o}s-R\'{e}nyi
hypergraphs with the Poisson degree distributions of the number of
triangles attached to vertices,
\begin{equation}
P_{3}(Q_{3})=\frac{(\langle Q_{3}\rangle)^{Q_{3}}}{Q_{3}!}e^{-\langle Q_{3} \rangle}.
\end{equation} \label{d ER-sgraph}

In the configuration model, the probability of the realization of a
given graph with a sequence of the number of triangles, $Q_{3}(1)$,
$Q_{3}(2)$, $Q_{3}(3)$,\dots,$Q_{3}(N)\equiv \{Q_{3}(i)\}$, attached
to vertices $i=1,2,\dots, N$ is defined by an equation,
\begin{equation}
G_{3}(\{a_{ijk}\})=\frac{1}{A}\prod_{i=1}^{N}\delta \Bigl(Q_{3}(i) -
\frac{1}{2}\sum_{j,k}a_{ijk}\Bigr)\prod_{i<j<k}p_{3}(a_{ijk}).
\label{s-config-model}
\end{equation}
Here $p_{3}(a_{ijk})$ is given by Eq.~(\ref{s-edge}). The
delta-function fixes the number of triangles attached to vertex $i$.
$A$ is a normalization constant. The distribution function of
triangles is
%%determined by the sequence $\{Q_{3}(i)\}$,
\begin{equation}
P_{3}(Q_{3})=\frac{1}{N}\sum_{i}\delta(Q_{3}-Q_{3}(i)).
\label{dd-t}
\end{equation}

%%%%%%%%%%%%%%%%%%%%%%%%%%%%%%%%%%%%%%%%%%%%%%%%%%%%%%%%%%%%%%%%%%%%
%%%%%%%%%%%%%%%%%%%%%%%%%%%%%%%%%%%%%%%%%%%%%%%%%%%%%%%%%%%%%%%%%%%%
%%%%%%%%%%%%%%%%%%%%%%%%%%%%%%%%%%%%%%%%%%%%%%%%%%%%%%%%%%%%%%%%%%%%
%%%%%%%%%%%%%%%%%%%%%%%%%%%%%%%%%%%%%%%%%%%%%%%%%%%%%%%%%%%%%%%%%%%%

One can further generalize equations Eqs.~(\ref{s-edge}) and
(\ref{p3}) and introduce the probability $p_{4}(a_{ijkl})$ that
vertices $i,j,k,$ and $l$ form a hyperedge of size 4 (a clique or
loop of size 4). In the case of loops, one can arrange these
vertices in order of increasing vertex index. Then, for a given
sequence of squares or cliques, $\{Q_{4}(1),Q_{4}(2), \dots\}$,
attached to vertices $i=1,2, \dots$, one can introduce the
probability $G_{4}(\{a_{ijkl}\})$ of the realization of a given
graph with a given sequence $\{Q_{4}(i)\}$ similar to
Eq.~(\ref{s-config-model}), and so on. The network ensemble of the
configuration model with a given sequences of edges $\{Q_{2}(i)\}$,
triangles $\{Q_{3}(i)\}$, squares $\{Q_{4}(i)\}$ and other motifs is
described by a product of the corresponding probabilities,
\begin{equation}
G(\{a_{ij}\},\{a_{ijk}\},\{a_{ijkl}\},...)=G_{2}(\{a_{ij}\})G_{3}(\{a_{ijk}\})...
\label{gs-config-model}
\end{equation}
The average of a quantity $K(\{a_{ij}\},\{a_{ijk}\},...)$ over the
network ensemble is
%%%%%%%%%%%%%%%%%%%%%%%%%%%%%
\begin{align}
&\langle K \rangle _{\text{en}} =\!\!\! \int K \Bigl(\{a_{ij}\},\{a_{ijk}\},\{a_{ijkl}\}...\Bigr) \times \nonumber \\
& G(\{a_{ij}\},\{a_{ijk}\},\{a_{ijkl}\},...) \prod_{i<j} da_{ij}  \!\!\!  \prod_{i<j<k} \!\!\! da_{ijk}  \!\!\!\!\!\! \prod_{i<j<k<l} \!\!\!\!\!\!  da_{ijkl}\dots,
\label{s-en-av}
\end{align}
where we integrate over all possible edges and hyperedges.

One can prove that the probability that different motifs have a
common edge, i.e., they are overlapping, tends to zero in the
infinite size limit $N\rightarrow \infty$. For example, the total
number of edges that overlap with triangles equals
\begin{equation}
\left\langle \frac{1}{2}\sum_{i,j,k}a_{ij}a_{ijk}\right\rangle
_{\text{en}}=\frac{N\left\langle Q_{2} \right\rangle\left\langle
Q_{3} \right\rangle}{N-1}.
\label{et}
\end{equation}
This number of double connections is finite in the limit
$N\rightarrow\infty$. Because the total number of edges is of the order
of $N$, the fraction of double connections tends to zero in the
thermodynamic limit. Thus, one can neglect overlapping among
different motifs. Using the standard methods in complex network
theory \cite{Bianconi:bc03,Bianconi:bm05a,Newman:n03b}, one can show
that in the configuration model the number of finite loops formed by
hyperedges is finite in the thermodynamic limit. Therefore, sparse
random uncorrelated hypergraphs have a hypertree-like structure.
%%However, a motif can have a strongly non-tree like  structure.

Uncorrelated random hypergraphs
%%in the configuration model
are described by a distribution function $P(Q_{2},Q_{3},Q_{4},\dots)$
which is the probability that a randomly chosen vertex has $Q_{2}$
edges, $Q_{3}$ triangles, $Q_{4}$ squares, and other motifs attached
to this vertex. In the case of uncorrelated motifs, we have
\begin{equation}
P(Q_{2},Q_{3},Q_{4},\dots)=\prod_{\ell=2}^{\infty}P_{\ell}(Q_{\ell}).
\label{dd}
\end{equation}
Here, $P_{\ell}(Q_{\ell})$ is the distribution function of loops of
size $\ell$,
\begin{equation}
P_{\ell}(Q_{\ell})=\frac{1}{N}\sum_{i}\delta(Q_{\ell}-Q_{\ell}(i)).
\label{l-dd-t}
\end{equation}
%%In other words, $P_{l}(Q_{l})$ is the probability that a randomly chosen
%%node have $Q_{l}$ motifs $l$ which are attached to this node.
In this kind of network, the number of nearest neighboring vertices
of vertex $i$ is equal to
\begin{equation}
Q(i)=Q_{2}(i) + 2Q_{3}(i)+2Q_{4}(i)+\dots \,\,. \label{degree}
\end{equation}
Therefore, the mean degree is
%%\begin{eqnarray}
%%\langle Q \rangle &=&\frac{1}{N} \sum_{i=1}^{N} Q(i) \nonumber \\[5pt]
%%&=&\sum_{Q_{2},Q_{3},\dots} Q P(Q_{2},Q_{3},Q_{4},\dots)
%%\label{m-degree}
%%\end{eqnarray}
%%%%%%%%%
\begin{equation}
\langle Q \rangle =\frac{1}{N} \sum_{i=1}^{N} Q(i)
=\sum_{Q_{2},Q_{3},\dots} Q P(Q_{2},Q_{3},Q_{4},\dots).
\label{m-degree}
\end{equation}

The local clustering coefficient $C(i)$ of node $i$ with degree
$Q(i)$, Eq.~(\ref{degree}), is determined by the number of triangles $Q_{3}(i)$
as follows,
\begin{equation}
C(i)=\frac{2Q_{3}(i)}{Q(i)(Q(i)-1)}.
\label{l-clustering1}
\end{equation}
Therefore, $C(i)$ is maximum if a network
%%under consideration
consists only  of triangles. In this case, we have $Q(i)=2Q_{3}(i)$
and therefore
\begin{equation}
C(i)=\frac{1}{Q(i)-1}.
\label{l-clustering2}
\end{equation}
On the other hand, if a network only consists of $\ell$-cliques with a given $\ell$ (fully
connected subgraphs of size $\ell$), then the local clustering
coefficient of vertex $i$ with $Q_{\ell}(i)$ attached cliques equals
\begin{equation}
C(i)=\frac{\ell-2}{Q(i)-1},
\label{l-clustering3}
\end{equation}
where $Q(i)=(\ell-1)Q_{\ell}(i)$ is degree of vertex $i$. If other motifs
(edges, squares, and larger finite loops) are present in the
network, then according to Eqs.~(\ref{l-clustering1}) and
(\ref{l-clustering3}) the local clustering coefficient $C(i)$ is
smaller than $(\ell-2)/[Q(i)-1]$. In Refs.
\cite{Serrano2006a,Serrano2006b,Serrano2006c}, it was shown that
properties of networks with "weak clustering", $C(Q) \sim O(1/Q)$, may
differ from properties of networks with "strong clustering" when
$C(Q)$ decreases slower than $1/Q$. The latter networks are beyond
the scope of the present article.

\section{Belief-propagation algorithm}
\label{algorithm}

Let us consider the Ising model with pairwise interactions between nearest neighboring spins on a sparse random network with
arbitrary distributions of motifs.
%%, which was introduced in the previous section.
The Hamiltonian of the model is
\begin{align}
E = &-\sum_{i}H_{i}S_{i}-\sum_{i<j}a_{ij}J_{ij}S_{i}S_{j} \nonumber \\
&-\sum_{i<j<k}a_{ijk}(J_{ij}S_{i}S_{j}+J_{jk}S_{j}S_{k}+J_{ik}S_{i}S_{k}) \nonumber \\
&- \!\!\! \sum_{i<j<k<l} \!\!\! a_{ijkl}(J_{ij}S_{i}S_{j}+J_{jk}S_{j}S_{k}+J_{kl}S_{k}S_{l}+\dots) \nonumber \\
%%&- \!\!\! \sum_{i<j<k<l} \!\!\! a_{ijkl}(J_{ij}S_{i}S_{j}+J_{jk}S_{j}S_{k}+J_{kl}S_{k}S_{l}+J_{li}S_{l}S_{i}) \nonumber \\
&+ ...~.
\label{Ising}
\end{align}
Here we sum the energies of spin clusters corresponding to motifs in the network. The second sum corresponds to simple edges. The third sum corresponds to triangles. The forth sum corresponds to motifs of size 4, and so on. $H_{i}$ is a
local magnetic field at vertex $i$.
%%interactions between spins ($S_{i}=\pm 1$) in motifs in a given network.
The coupling $J_{ij}$ determines the energy of
pairwise interaction between spins $i$ and $j$. In general, one can introduce multi-spin interactions between spins in hyperedges, for example, $S_{i}S_{j}S_{k}$, $S_{i}S_{j}S_{k}S_{l}$, and so on. However, this kind of interaction is beyond the scope of our article. Generally, the local fields $H_{i}$ and
the couplings $J_{ij}$ can be random quantities.

\begin{figure}[t]
\begin{center}
\scalebox{0.5}{\includegraphics[angle=0]{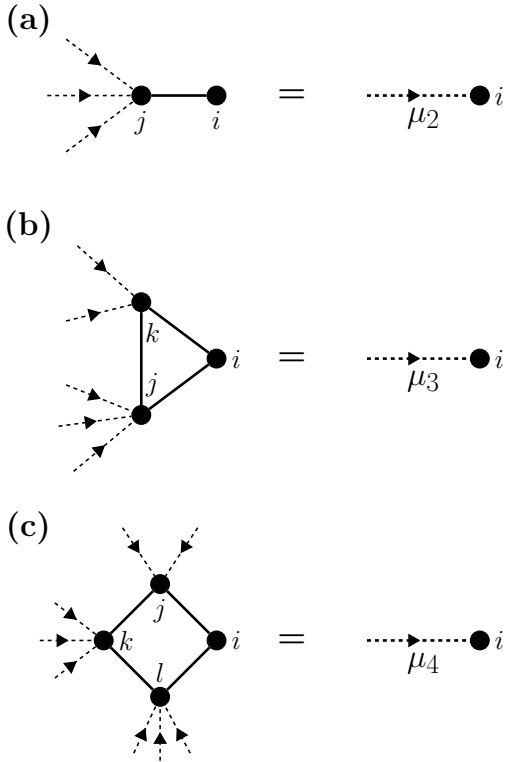}}
%%\scalebox{0.4}{\includegraphics[angle=0]{test_fig.eps}}
\end{center}
\caption{ Diagram representation of the belief-propagation update
rules for messages from a motif to a destination vertex $i$. Arrows
represent incoming messages. [(a), (b) and (c)] Update rules for an edge, a triangle, and a square. Mathematical
expressions of these diagrams are given by Eqs.~(\ref{ur}) and
(\ref{hyper-ur}).} \label{fig-BP}
\end{figure}

In order to solve this model, we generalize the belief-propagation
algorithm. At first, we note the belief-propagation algorithm in
application for the Ising model on tree-like uncorrelated complex
networks without short loops (for more details, see Ref.
\cite{Dorogovtsev:dgm08}). In this case, there are only edges
determined by the adjacency matrix $a_{ij}$ and the clustering
coefficient is zero in the thermodynamic limit. Consider spin $i$.
A nearest neighboring spin $j$ sends  a so-called {\em
message} to spin $i$ that we denote as $\mu _{ji}(S_{i})$. This message is
normalized,
\begin{equation}
\sum_{S_{i}=\pm 1}\mu _{ji}(S_{i})=1.
\label{message-n}
\end{equation}
%%It has a probabilistic interpretation (see Ref.~\cite{Dorogovtsev:dgm08}).
Within the belief-propagation
algorithm, the probability that spin $i$ is in a state $S_{i}$ is
determined by the normalized product of incoming messages to spin
$i$ and the probabilistic factor $e^{\beta H_{i}S_{i}}$ due to a
local field $H_{i}$:
\begin{equation}
p_{i}(S_{i})=\frac{1}{A}e^{\beta H_{i}S_{i}}\prod_{j}\mu
_{ji}(S_{i}), \label{p}
\end{equation}
where $A$ is a normalization constant and $\beta=1/T$ is the reciprocal temperature. Now we can calculate the mean
moment of spin $i$,
\begin{equation}
\langle S_{i}\rangle= \sum_{S_{i}=\pm 1}S_{i} p_{i}(S_{i}).
\label{mean spin1}
\end{equation}
A message $\mu _{ji}(S_{i})$ can be written in a general form,
\begin{equation}
\mu _{ji}(S_{i})=\exp(\beta h_{ji}S_{i})/[2\cosh(\beta h_{ji})],
\label{message-r}
\end{equation}
Using this
representation, we rewrite Eq.~(\ref{mean spin1}) in a physically
clear form,
\begin{equation}
\langle S_{i}\rangle= \tanh\Bigl( \beta H_{i} + \beta \sum_{j} a_{ji} h_{ji} \Bigr).
\label{mean spin 2}
\end{equation}
This equation shows that a parameter $h_{ji}$ plays the role of an
effective field produced by spin $j$ at site $i$. Messages $\mu
_{ji}(S_{i})$ obey a self-consistent equation that is called
\emph{update rule},
\begin{equation}
B\sum_{S_{j}=\pm 1}e^{- \beta E(j,i)}\prod_{n \neq i}\mu
_{nj}(S_{j})=\mu _{ji}(S_{i}), \label{ur}
\end{equation}
where the index $n$ numerates nearest neighbors of vertex $j$ and $B$
is a normalization constant. The diagram representation of this
equation is shown in Fig.~\ref{fig-BP} $(a)$. The probabilistic
factor $\exp[-\beta E(j,i)]$ takes account of the interaction energy of
spins $j$ and $i$ and a local field $H_{j}$,
\begin{equation}
E(j,i)= -H_{j}S_{j} - a_{ji}J_{ji} S_{i} S_{j}.
\label{delta-E}
\end{equation}
Thus, a message from $j$ to $i$ is determined by the coupling
between these spins and messages received by $j$ from its neighbors
except $i$. Multiplying Eq.~(\ref{ur}) by $S_{i}$ and summing over
$S_{i}=\pm 1$, we obtain a self-consistent equation for the
effective fields $h_{ji}$,
\begin{equation}
\tanh (\beta h_{ji})=\tanh \Bigl(\beta a_{ji}J_{ji}\Bigr) \tanh \Bigl[\beta (H_{j}+\!\sum_{n (\neq i)}\!a_{nj}h_{nj})\Bigr].
\label{recursion}
\end{equation}
In the thermodynamic limit, this equation is exact for a tree-like
graph. The Bethe-Peierls approach and the Baxter recurrence method
give exactly the same result \cite{Dorogovtsev:dgm08,bbook82}.
%%. For more details see Refs. \cite{Dorogovtsev:dgm08,bbook82} and references therein.
In numerical calculations, Eqs.~(\ref{ur}) and (\ref{recursion}) may
be solved by use of iterations. In a general case, an
analytical solution of these equations is unknown.
%%except for some cases see \cite{Dorogovtsev:dgm08}.
In the case of all-to-all interactions with random couplings
$J_{ji}$ (the Sherrington-Kirkpatrick model), this set of equations
is reduced to well-known TAP equations \cite{Thouless:tap77} that
are exact in the thermodynamic limit. These equations also give an
exact solution of the ferromagnetic Ising model with uniform
couplings $J_{ji}=J>0$ on a random uncorrelated graph with zero
clustering coefficient \cite{dgm02,lvvz02,Dorogovtsev:dgm08}.

%%%%%%%%%%%%%%%%%%%%%%%%%%%%%%%%%%%%%%%%%%%%%%%%%%%%%%%%%%%%%%%%%%%%
%%%%%%%%%%%%%%%%%%%%%%%%%%%%%%%%%%%%%%%%%%%%%%%%%%%%%%%%%%%%%%%%%%%%
%%%%%%%%%%%%%%%%%%%%%%%%%%%%%%%%%%%%%%%%%%%%%%%%%%%%%%%%%%%%%%%%%%%%
%%%%%%%%%%%%%%%%%%%%%%%%%%%%%%%%%%%%%%%%%%%%%%%%%%%%%%%%%%%%%%%%%%%%

We now generalize the belief-propagation algorithm to the case of
networks with given distributions of motifs described in Sec.
\ref{config}. In a network that consists only of edges, a message
goes along an edge from the spin at one edge end to the spin at the
other edge end. For the networks with motifs, it is
natural to introduce a message that is sent by a motif attached to
a vertex. This message goes to a spin to which this motif
(hyperedge) is attached. Different motifs may be attached to a
vertex, see Fig.~\ref{fig1}.  Let $M_{\ell}(i)$ denote a cluster of
size $\ell$ attached to vertex $i$ that we will call
the \emph{destination vertex}. Vertices $j_{1},j_{2},...,j_{\ell-1}$
together with the destination vertex $i$ form this motif. We
introduce a message $\mu _{M_{\ell}(i)}(S_{i})$ to spin $i$ from a
motif $M_{\ell}(i)$.
%%, i.e., from the spins $j_{1},j_{2},...j_{\ell-1}$ forming this motifs.
This message is
normalized and can be written as follows,
\begin{equation}
\mu_{M_{\ell}(i)}(S_{i})=\exp\Bigl(\beta
h_{M_{\ell}(i)}S_{i}\Bigr)/[2\cosh(\beta h_{M_{\ell}(i)})].
\label{hyper-a-message}
\end{equation}
As above, the probability that spin $i$ is in a state $S_{i}$ is
determined by the normalized product of incoming messages from
motifs attached to spin $i$ and the probabilistic factor $e^{\beta
H_{i}S_{i}}$,
\begin{equation}
p_{i}(S_{i})=\frac{1}{A}e^{\beta H_{i}S_{i}}
\prod_{\{M_{\ell}(i)\}}\mu_{M_{\ell}(i)}(S_{i}). \label{message-M}
\end{equation}
Using Eq.~(\ref{mean spin1}), we find that the mean moment $\langle
S_{i}\rangle$
%%of spin $i$
is determined by effective fields $h_{M_{\ell}(i)}$ acting on spin
$i$ from motifs $M_{\ell}(i)$, $\ell=2,3,4,...$, attached to
$i$, see Fig.~\ref{fig1},
\begin{equation}
\langle S_{i}\rangle= \tanh\Bigl( \beta H_{i} + \beta
\sum_{\{M_{\ell}(i)\}} h_{M_{\ell}(i)} \Bigr). \label{mean spin 2}
\end{equation}
Here the sum is taken over motifs attached to $i$.

Let us find an update rule for a message $\mu_{M_{\ell}(i)}(S_{i})$
from a given motif $M_{\ell}(i)$ to vertex $i$. We introduce the
following update rule:
\begin{equation}
B\!\! \!\!\sum_{\{S_{j}=\pm 1 \}}\!\! \!\!e^{- \beta E(M_{\ell}(i))}\prod_{j
} \!\prod_{\{M_{n}(j)\neq M_{\ell}(i)\}}\!\!\!\!\!\!\!\!\!\!\!\mu_{M_{n}(j)}(S_{j})=\!\mu_{M_{\ell}(i)}(S_{i}).
\label{hyper-ur}
\end{equation}
This rule shows that the message $\mu_{M_{\ell}(i)}(S_{i})$ is
equal to the product of incoming messages from motifs attached to all spins $j$ in the motif $M_{\ell}(i)$ except spin $i$ and the motif $M_{\ell}(i)$ itself
%%to the spins $j_{1},j_{2},...$ in this motif $M_{\ell}(i)$ except a
%%message from the motif $M_{\ell}(i)$.
(see
%%Diagram representations of this update rule for simple motifs are shown in
Fig.~\ref{fig-BP}.
In Eq.~(\ref{hyper-ur}) we average over all spin configurations of
the spins $S_{j_{1}},S_{j_{2}},...$, and $S_{j_{\ell-1}}$ in the
motif. $B$ is a normalization constant. $E[M_{\ell}(i)]$ is an
energy of the interaction between spins in the motif $M_{\ell}(i)$.
This update rule also takes account of local fields $H_j$ acting on all spins
except the destination spin $i$. We note that Eq.~(\ref{hyper-ur}) is valid for arbitrary motifs even with a complex
internal structure. The only assumption is that a sparse network
under consideration, in terms of hypergraphs, has a hypertree-like
structure. Messages can be found numerically by use of iterations of Eq.~(\ref{hyper-ur}) that start from an initial distribution of the messages.

\begin{figure}[t]
\begin{center}
\scalebox{0.5}{\includegraphics[angle=0]{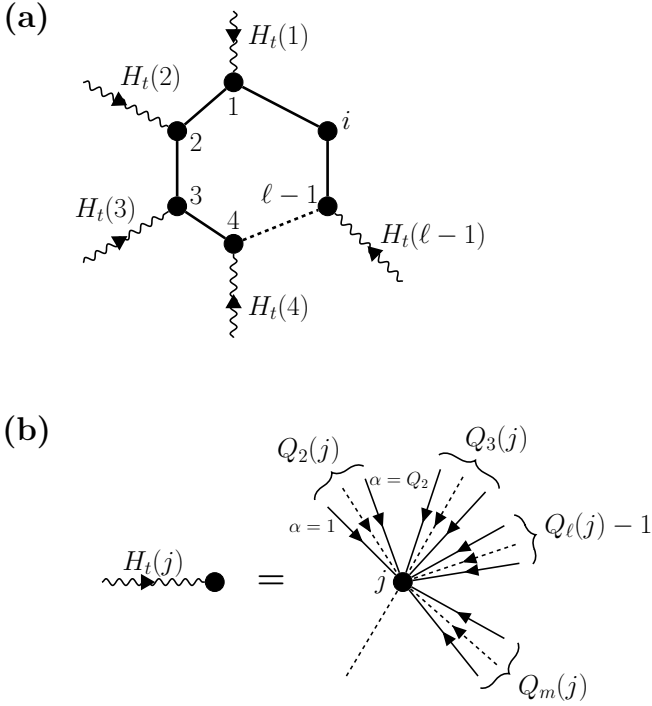}}
\end{center}
\caption{ (a) A loop (motif $M_{\ell}$) of size $\ell$ is attached
to destination vertex $i$. Wave arrows represent total effective
fields $H_{t}(j)$ acting on spins with index $j=1,2, \dots,
\ell-1$ in this loop.  (b) In turn, the total effective field
$H_{t}(j)$ is a sum of effective fields produced by all motifs
attached to vertex $j$ except the motif $M_{\ell}$. There are
$Q_{2}(j)$ edges, $Q_{3}(j)$ triangles, and so on. However, the
number of loops of size $\ell$ equals $Q_{\ell}(j)-1$, because we
should not account the motif $M_{\ell}$ that is common for vertices
$j$ and $i$.} \label{fig3}
\end{figure}

%%Let us consider a simple case when
For the sake of simplicity, let motifs $M_{\ell}(i)$ be finite
loops of size $\ell=2,3,\dots$.  Then
\begin{equation}
E(M_{\ell}(i))= - \sum_{n=1}^{\ell-1} H_{j_{n}}S_{j_{n}} -
\sum_{n=0}^{\ell-1} J_{j_{n}j_{n+1}}S_{j_{n}}S_{j_{n+1}},
\label{E0-Mi}
\end{equation}
where $j_{0}=j_{\ell}\equiv i$. If motifs are $\ell$-cliques, then the energy $E(M_{\ell}(i))$ takes account of interactions between all spins in these cliques,
\begin{equation}
E(M_{\ell}(i))= - \sum_{j(\neq i)} H_{j} S_{j}
- \frac{1}{2}\sum_{i,j\in M_{\ell}(i)} J_{i,j}S_{i}S_{j} \text{.} \label{E-Mi-clique}
\end{equation}
Multiplying Eq.~(\ref{hyper-ur}) by
$S_{i}$ and summing over all spin configurations, we obtain an
equation for the effective field $h_{M_{\ell}(i)}$,
\begin{equation}
\tanh\Bigl( \beta h_{M_{\ell}(i)}\Bigr)=\langle S_{i}\rangle_{M_{\ell}(i)}.
%%F_{M(i)}[\{H_{j}\},\{H_{eff}(j)\}]
\label{recursion-M}
\end{equation}
The function on the right hand side is
\begin{equation}
\langle S_{i}\rangle_{M_{\ell}(i)} =
\frac{1}{Z(M_{\ell}(i))}\sum_{\{S_{i},S_{j_{1}}, ... =\pm
1\}}\!\!\!\!S_{i}e^{- \beta \widetilde{E}(M_{\ell}(i))} \text{.}
\label{FM}
\end{equation}
This function has a meaning of the mean moment of spin $S_i$ in the cluster $M_{\ell}(i)$
with an energy
\begin{equation}
\widetilde{E}(M_{\ell}(i))= - \sum_{n=1}^{\ell-1} H_{t}(j_{n})S_{j_{n}}
- \sum_{n=0}^{\ell} J_{j_{n}j_{n+1}}S_{j_{n}}S_{j_{n+1}} \text{.} \label{E-Mi}
\end{equation}
%%%%%%%%%%
This energy takes into account both the interaction between spins in this motif
and total effective fields $H_{t}(j)$ acting on these spins. In turn, the field $H_{t}(j)$
acting on $j$ is the sum of a local field $H_{j}$ and effective
fields $h_{M_{m}(j)}$ produced by incoming messages from motifs
$M_{m}(j)$ attached to vertex $j$ except the motif $M_{\ell}(i)$,
see Fig.~\ref{fig3},
\begin{equation}
H_{t}(j)=H_{j}  + \sum_{\{M_{m}(j)(\neq M_{\ell}(i))\}}h_{M_{m}(j)}.
\label{Ht}
\end{equation}
$Z[M_{\ell}(i)]$ is the partition function of the cluster $M_{\ell}(i)$,
\begin{equation}
Z[M_{\ell}(i)]=\sum_{\{S_{i},S_{j_{1}},...=\pm 1\}}e^{- \beta \widetilde{E}(M_{\ell}(i))}.
\label{ZM}
\end{equation}
Thus, the function $\langle S_{i}\rangle_{M_{\ell}(i)}$ is a
function $F [H_{t}(j_{1}),H_{t}(j_{2}),\dots,H_{t}(j_{\ell -1})]$ of
the total fields $H_{t}(j)$ acting on all spins in the motif
$M_{\ell}(i)$ except spin $i$. For a given network with motifs, it
is necessary to solve Eq.~(\ref{recursion-M}) with respect to the
effective fields $h_{M_{\ell}(i)}$. One then can calculate the mean
magnetic moments $\langle S_{i}\rangle$ from Eq.~(\ref{mean spin
2}). Note that the only condition we used to derive the equations
above was hypertree-like structure of the networks. The absence of
correlations in these hypertree-like networks was not needed.
Equations~(\ref{hyper-ur}) and ~(\ref{recursion-M}) are our main
result that actually generalizes the Bethe-Peierls approach and the
Baxter recurrence method.

Let us study the ferromagnetic Ising model with uniform couplings,
$J_{ji}=J>0$, at zero magnetic field, i.e., $H_{i}=H=0$, on a
uncorrelated hypergraph which consists of edges and finite loops.  In
a random network, effective fields $h_{M_{\ell}(i)}$  are also
random variable and fluctuate from vertex to vertex. For a given
$\ell$, we introduce the distribution function of fields
$h_{M_{\ell}(i)}$,
\begin{equation}
\Psi_{\ell}(h)=\frac{1}{A}
\sum_{i=1}^{N}\sum_{\{M_{\ell}(i)\}}\delta(h-h_{M_{\ell}(i)}). \label{psi}
\end{equation}
Here, we sum over vertices $i$ and attached motifs $M_{\ell}(i)$ of
size $\ell$. $A=N\langle Q_{\ell} \rangle$ is the normalization
constant. We assume that in the thermodynamic limit, $N\rightarrow
\infty$, the average over vertices in the network is equivalent to
the average over the statistical network ensemble,
Eq.~(\ref{s-en-av}). Using Eq.~(\ref{recursion-M}), we obtain
an equation for the distribution function $\Psi_{\ell}(h)$,
%%%
\begin{equation}
\Psi_{\ell}(h)=\int \delta \Bigl(h-T\tanh ^{-1}[\langle S \rangle_{M_{\ell}}] \Bigr)\prod_{j=1}^{\ell-1} \Phi_{\ell}(H_{t}(j)) d H_{t}(j).
\label{psi}
\end{equation}
Here $\langle S \rangle_{M_{\ell}}$ is the function $F
[H_{t}(1),H_{t}(2),\dots,H_{t}(\ell -1)]$ defined by Eq.~(\ref{FM}).
$H_{t}(j)$ is a total field, Eq.~(\ref{Ht}),  acting on vertex
$j=1,2, \dots,\ell-1$ in the motif $M_{\ell}$, see Fig.~\ref{fig3}.
The integration is over fields $H_{t}(j)$ with the
distribution function $\Phi_{\ell}(H_{t}(j))$. Using Eq.~(\ref{Ht}),
we can find this function,
\begin{align}
&\Phi_{\ell}(H) = \sum_{Q_{2},Q_{3},...} \!\!\! P(Q_{2},Q_{3},...)\frac{Q_{\ell}}{\langle Q_{\ell}\rangle}\times   \nonumber \\
&\int \delta \Bigl(H - \!\!\! \sum_{m (\neq \ell)}^{\infty}
\sum_{\alpha=1}^{Q_m} h_{M_{m}}(\alpha) - \!\!\!
\sum_{\alpha=1}^{Q_{\ell}-1} h_{M_{\ell}}(\alpha)\Bigr) \times
\nonumber \\
&\prod_{m (\neq \ell)}^{\infty} \!\! \Bigl(
\prod_{\alpha=1}^{Q_{m}}\!\!\Psi_{m}(h_{M_{m}}(\alpha))dh_{M_{m}}(\alpha)
\Bigr)
\!\!\!\! \prod_{\alpha=1}^{Q_{\ell}-1} \! \! \Psi_{\ell}(h_{M_{\ell}}(\alpha)) dh_{M_{\ell}}(\alpha). \nonumber \\
\label{phi}
\end{align}
Here $P(Q_{2},Q_{3},...)$ is given by Eq.~(\ref{dd}). This is the
probability that a destination vertex has $Q_{2}$ edges, $Q_{3}$
triangles, and so on. In turn, incoming messages $h_{M_{m}}(\alpha)$
to the destination spin from attached loops of size $m$ are
numerated by the index $\alpha$, $\alpha=1,2, ..., Q_{m}$. Only for
incoming messages from loops of size $\ell$ do we have $\alpha=1,2,
..., Q_{\ell}-1$, because we should not account for the motif
$M_{\ell}$. The integration is over incoming messages to all
vertices in the motif (hyperedge) $M_{\ell}$ except the destination
vertex $i$, see Fig.~\ref{fig3}.  Note that we have an additional
factor $Q_{\ell}/\langle Q_{\ell}\rangle$, because there is the probability in the
configuration model  $P_{\ell}(Q_{\ell})Q_{\ell}/\langle
Q_{\ell}\rangle$  that if we arrive at a
destination vertex along a hyperedge $M_{\ell}$, then this vertex
has $Q_{\ell}-1$ outgoing hyperedges $M_{\ell}$. Equations
~(\ref{psi}) and ~(\ref{phi}) represent a set of self-consistent
equations for the functions $\Psi_{\ell}(h)$,
$\ell=2,3,4, \dots \,$. Using Eq.~(\ref{mean spin 2}), one then can
find a distribution function of spontaneous magnetic moments
$\langle  S_i \rangle$ in the network. Equations
~(\ref{psi}) and ~(\ref{phi}) are also valid for networks with cliques.
%%%%%%%%%%

%%%%%%%%%%%%%%%%%%%%%%%%%%%%%%%%%%%%%%%%%%%%%%%%%%%%%%%%%%%%%%%%%%%%
%%%%%%%%%%%%%%%%%%%%%%%%%%%%%%%%%%%%%%%%%%%%%%%%%%%%%%%%%%%%%%%%%%%%
%%%%%%%%%%%%%%%%%%%%%%%%%%%%%%%%%%%%%%%%%%%%%%%%%%%%%%%%%%%%%%%%%%%%
%%%%%%%%%%%%%%%%%%%%%%%%%%%%%%%%%%%%%%%%%%%%%%%%%%%%%%%%%%%%%%%%%%%%

It should be noted that our approach is similar to the generalized belief-propagation algorithm proposed in Ref.~\cite{Yedidia:yfw01}. It was shown that the belief-propagation algorithm is equivalent to the cluster variation method (CVM) introduced by R.~Kikuchi (see Ref.~\cite{Pelizzola05} for more details).
Note that in the model of complex networks in the present work, two motifs can only overlap each other by a single node. They have no joint links.
%%The important peculiarity of considered complex networks with motifs is that the motifs only overlap each other at one node to which they are attached.
%%Moreover, motifs
Motifs here play the role of ``hyperedges'' in contrast to the approach in Ref.~\cite{Yedidia:yfw01} where clusters are considered as ``super-nodes''. Finally, we assumed that networks under consideration have a hypertree-like structure in the infinite size limit.

\section{Critical temperature and critical behavior}
\label{crit}

\begin{figure}[t]
\begin{center}
\scalebox{0.7}{\includegraphics[angle=0]{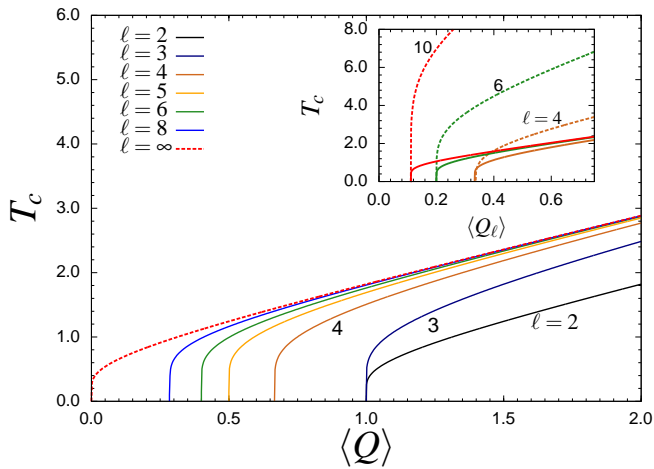}}
\end{center}
\caption{ (Color online) Critical temperature $T_c$ of the
ferromagnetic Ising model versus the mean number of nearest
neighbors $\langle Q \rangle =2\langle Q_{\ell}\rangle$ in
uncorrelated random networks with Poisson distribution of loops of
size $\ell$ and the mean number $\langle Q_{\ell}\rangle$ of these
loops attached to a vertex. Note that the case $\ell=2$ is special:
there are no loops, and we have the Erd\H{o}s-R\'{e}nyi graph. Inset
displays $T_c$ versus the mean number $\langle Q_{\ell}\rangle$ of
$\ell$-cliques (dashed lines), and, for comparison, $T_c$ for
finite loops of the same size $\ell$ (solid lines). Here we set $J=1$.} \label{figTc}
\end{figure}

In the paramagnetic phase at zero magnetic field, there is no
spontaneous magnetization and the effective fields are equal to zero, i.e.,
$h_{M_{\ell}}=0$. Therefore, we obtain
\begin{equation}
\Psi_{\ell}(h)=\delta(h).
\label{psi0}
\end{equation}
for any $\ell$. This is the only solution of Eqs.~(\ref{psi}) and
~(\ref{phi}) in the paramagnetic phase. Below a critical temperature
$T_c$, this solution becomes unstable and a non-trivial solution for
the function $\Psi_{\ell}(h)$ emerges. In this case, a mean value of
the effective field $h_{M_{\ell}}$,
\begin{equation}
\langle h_{M_{\ell}} \rangle_{T} = \int h \Psi_{\ell}(h)dh,
\label{mean -h}
\end{equation}
becomes non-zero. Here $\langle ... \rangle_{T}$ denotes the
thermodynamic average. In the case of a continuous phase transition,
$\langle h_{M_{\ell}} \rangle_{T}$ tends to zero if the temperature $T$
tends to $T_c$ from below. Let use this fact. From
Eq.~(\ref{recursion-M}) we obtain
%%%
\begin{equation}
\int \tanh (\beta h) \Psi_{\ell}(h)dh =\int \langle S \rangle_{M_{\ell}} \prod_{j=1}^{\ell-1} \Phi_{\ell}(H_{t}(j)) d H_{t}(j).
\label{psi2}
\end{equation}
%%%
We expand the functions $\tanh (\beta h)$ and $\langle S
\rangle_{M_{\ell}}$ on the left and right hand sides of this
equation in small $h$ and $H_{t}(j)$, respectively:
\begin{eqnarray}
\tanh (\beta h)=\beta h + O(h^3),
\nonumber
\\[5pt]
\langle S_{i} \rangle_{M_{\ell}}=\sum_{j=1}^{\ell-1} \chi_{\ell}(ij)
H_{t}(j)+ O(H_{t}^{3}(j)), \label{exp1}
\end{eqnarray}
where we define
\begin{equation}
\chi_{\ell}(ij)\equiv \frac{\partial \langle S_{i}
\rangle_{M_{\ell}}}{\partial H_{t}(j)}\Big |_{H_{t}(j)=0}=\beta
\langle S_{i}S_{j} \rangle_{M_{\ell}} \Big|_{H_{t}(j)=0}.
\label{exp2}
\end{equation}
$\chi_{\ell}(ij)$ is a non-local susceptibility in a spin cluster
${M_{\ell}}$ at zero magnetic field. Note that $T
\chi_{\ell}(ii)=1$. Assuming that at $T \rightarrow T_{c}-0$ the
first moment of the function $\Psi_{\ell}(h)$, i.e., $\langle
h_{M_{m}} \rangle_{T}$, is much larger than higher moments, i.e.,
$\langle h_{M_{m}} \rangle_{T} \gg \langle h_{M_{m}}^{n}
\rangle_{T}$, in the leading order we obtain a linear equation,
\begin{eqnarray}
\langle h_{M_{\ell}} \rangle_{T}= \frac{F_{\ell}(T)}{\langle Q_{\ell} \rangle} \Bigl[\langle Q_{\ell}(Q_{\ell}-1) \rangle \langle h_{M_{\ell}} \rangle_{T} \nonumber
\\[5pt]
+ \sum_{m(\neq \ell)}^{\infty} \langle Q_{\ell}Q_{m} \rangle \langle
h_{M_{m}} \rangle_{T} \Bigr ], \label{exp3}
\end{eqnarray}
where $\langle \dots \rangle$ defines an average over the
distribution function $P(Q_{2},Q_{3},Q_{4},\dots)$ given by Eq.
(\ref{dd}). The function $F_{\ell}(T)$ is defined as follows,
%%%%%
\begin{equation}
F_{\ell}(T)\equiv T\sum_{j=1}^{\ell-1}\chi_{\ell}(ij)=T\chi_{\ell}-1
, \label{Fl}
\end{equation}
$\chi_{\ell}$ is the total zero-field magnetic susceptibility of the Ising model on a ring of size $\ell$.
Simple calculations give
%%%%%%
\begin{eqnarray}
F_{2}(T)=t, \,\,\,\,\,\,\,\,\,\, \ell=2, \nonumber
\\[5pt]
F_{\ell}(T)=\frac{2t(1-t^{\ell-1})}{(1-t)(1+t^{\ell})}, \,\,\,\,\,\,\,\,\,\,
\ell\geq 3,\label{F}
\end{eqnarray}
where $t=\tanh (J/T)$. In the paramagnetic phase, the set of linear
equations (\ref{exp3}) for parameters $\langle h_{M_{\ell}}
\rangle_{T}$ has only a trivial solution, $\langle h_{M_{\ell}}
\rangle_{T}=0$.  A non-trivial solution appears at a temperature at
which
\begin{equation}
det\widehat{M}=0, \label{Tc0}
\end{equation}
where the matrix $M_{mn}$ is defined as follows:
\begin{eqnarray}
M_{mn}=\langle Q_{m}Q_{n}\rangle, \,\,\,\,\,\,\,\,\,\,   m\neq n,
\nonumber
\\[5pt]
M_{mm}=\langle Q_{m}(Q_{m}-1)\rangle - \frac{\langle
Q_{m}\rangle}{F_{m}(T)},\label{Psi-ei}
\end{eqnarray}
for $m,n=2,3,...\,$. Equation (\ref{Tc0}) determines the critical
point $T_c$ of the continuous phase transition. Below $T_c$,
spontaneous effective fields appear, i.e., $\langle h_{M_{\ell}}
\rangle_{T} \neq 0$.  Therefore, there is a non-zero spontaneous
magnetic moment. If all motifs are loops of equal length $\ell$,
then the critical temperature $T_c$ is determined by the equation
\begin{equation}
B_{\ell}F_{\ell}(T)=1
\label{Tc1}
\end{equation}
where $B_{\ell}$ is the average branching coefficient for these hyperedges ($\ell$-loops in this network),
\begin{equation}
B_{\ell}=\frac{\langle Q_{\ell}(Q_{\ell}-1)\rangle}{\langle Q_{\ell}\rangle} .
\label{B}
\end{equation}
In particular, if there are only edges, i.e., $\ell=2$, then Eq. (\ref{Tc1})
gives
\begin{equation}
T_c=2J/\ln \Bigl (\frac{B_{2}+1}{B_{2}-1} \Bigr). \label{Tc3}
\end{equation}
This result was found for uncorrelated random complex networks with
arbitrary degree distributions
\cite{dgm02,lvvz02,Dorogovtsev:dgm08}.
In the limit $B_{\ell} \gg 1$,
we find that
\begin{eqnarray}
T_{c}(\ell=2)/J \cong B_{\ell} + o(1), \label{asympt-Tc2}
\\[5pt]
T_{c}(\ell \geq 3)/J \cong 2B_{\ell} +
1+o(1),\label{asympt-Tc}
\end{eqnarray}
Figure~\ref{figTc} shows the
dependence of the critical temperature $T_{c}(\ell)$ on the mean number
of nearest neighbors $\langle Q \rangle$ in the networks with the Poissonian distribution of $\ell$-loop motifs.
%%, in which the Poissonian distribution of hyperdegrees is assumed.
In this case the average branching coefficient is $B_{\ell}=Q_{\ell}$. Notice also that at $\ell \geq 3$ we have  $\langle Q \rangle =2\langle Q_{\ell}\rangle=2B_{\ell}$ according to Eqs.~(\ref{degree}) and (\ref{m-degree}).
%%In this case $B_{\ell}=Q_{\ell}$.
%%One can see that, at a given $\langle
%%Q \rangle $, the larger the size of loops the larger the critical temperature.
%%At $\ell=2$ (an ordinary Erd\H{o}s-R\'{e}nyi graph),
%%the Ising model has a lower $T_c$ in comparison to that at $\ell \geq 3$.
Comparison between Eqs.~(\ref{asympt-Tc2}) and (\ref{asympt-Tc}) shows that at a given
%%average branching coefficient $B_{\ell} \gg 1$
mean degree $Q \gg 1$ for the Poisson distribution of hyperdegrees
the critical temperature $T_{c}(\ell >2)$ is higher than $T_c(\ell=2)$ only by 1. This shift is the only effect of finite loops.
%% in networks without loops.

If there are only two motifs, for example, loops of size $\ell$ and
$\ell'$, then Eq. (\ref{Tc0}) leads to the equation
\begin{equation}
\Bigl[ B_{\ell} - \frac{1}{F_{\ell}(T)}\Bigr]\Bigl[ B_{\ell'} -
\frac{1}{F_{\ell'}(T)} \Bigr]=\frac{\langle
Q_{\ell}Q_{\ell'}\rangle^2}{\langle Q_{\ell}\rangle \langle
Q_{\ell'}\rangle}. \label{Tc4}
\end{equation}

%%%%%%%%%%%%%%%%%%%%%%%%%%%%%%%%%%%%%%%%%%%%%%%%%%%%%%%%%%%%%%%%%%%%
%%%%%%%%%%%%%%%%%%%%%%%%%%%%%%%%%%%%%%%%%%%%%%%%%%%%%%%%%%%%%%%%%%%%
%%%%%%%%%%%%%%%%%%%%%%%%%%%%%%%%%%%%%%%%%%%%%%%%%%%%%%%%%%%%%%%%%%%%
%%%%%%%%%%%%%%%%%%%%%%%%%%%%%%%%%%%%%%%%%%%%%%%%%%%%%%%%%%%%%%%%%%%%
From Eqs.~(\ref{F}) and (\ref{Tc1}), it follows that if an
uncorrelated random hypergraph has a divergent second moment
$\langle Q_{m}^2 \rangle$ for any $m \geq 2$, then $T_c$ becomes
infinite in the infinite size limit. This result was obtained for
ordinary uncorrelated random complex network in Refs.
\cite{dgm02,lvvz02,Dorogovtsev:dgm08}.
%%Moreover, because equations
%%(\ref{psi}) and (\ref{phi}) have analytical properties similar to
%%ones studied in Refs. \cite{dgm02,lvvz02,Dorogovtsev:dgm08}, we
%%conclude that the ferromagnetic Ising model on considered loopy
%%networks has the same critical properties as ones on ordinary
%%uncorrelated random networks.

Another example of complex motifs are
%%a fully connected subgraph of $\ell$ verticies, or, in other words, this is a
$\ell$-cliques. In this kind of complex networks the clustering coefficient is larger than in networks with loops of size $\ell=3$ (compare between Eqs.~(\ref{l-clustering2}) and (\ref{l-clustering3})). Calculating the susceptibility $\chi$ of a cluster
of spins on this subgraph, we can find the function
$F_{\ell}(T)$ in Eq.~(\ref{Fl}) at $\ell \geq 3$,
\begin{equation}
F_{\ell}(T)=(\ell-1) \Bigl[1-\frac{4 I_{\ell-2}(T)}{I_{\ell}(T)}
\Bigr]. \label{l-clique1}
\end{equation}
Here we introduced the function
\begin{equation}
I_{\ell}(T)=\sum_{n=0}^{\ell}C_{n}^{\ell}\exp\Bigr[\frac{1}{2}\beta
J (2n - \ell)^2 \Bigl],
 \label{l-clique2}
\end{equation}
where $C_{n}^{\ell}=n!/(n-\ell)!\ell!$ is the binomial coefficient. If a
network only consists of uncorrelated $\ell$-cliques then the
critical temperature of the ferromagnetic Ising model is given by
Eq.~(\ref{Tc1}) with the function Eq.~(\ref{l-clique1}).
%%In the case
%%of the Poisson distribution of $\ell$-cliques with the mean degree
%%$\langle Q_{\ell} \rangle \gg 1$,
In the limit $B_{\ell} \gg 1$ we obtain an asymptotic result,
\begin{equation}
T_{c}\approx (\ell-1)B_{\ell}+(\ell-2)(1+o(1)).
 \label{Tc-clique}
\end{equation}
Let us consider the Poisson distribution. In this case we have $B_{\ell}=\langle Q_{\ell} \rangle$ and the mean degree is $\langle Q \rangle = (\ell - 1) \langle Q_{\ell} \rangle=(\ell - 1) B_{\ell}$. Critical temperatures of networks with $\ell-$cliques and $\ell-$loops are compared in Fig.~\ref{figTc}.
Comparison between Eqs.~(\ref{Tc-clique}), (\ref{asympt-Tc2}) and (\ref{asympt-Tc}) shows that at a given
%%average branching coefficient $B_{\ell} \gg 1$
mean degree $\langle Q \rangle \gg 1$
%%for the Poisson distribution of hyperdegrees
the critical temperature $T_{c}(\ell)$ in Eq.~(\ref{Tc-clique}) is larger than the critical temperature Eq.~(\ref{asympt-Tc2}) only by $\ell-2$. This shift is the only effect of clustering. It gives a relative correction of order $O[(\ell-2)/\langle Q \rangle]$, because $T_c (\ell)/T_c(\ell =2)\approx 1+(\ell-2)/\langle Q \rangle $.
This result shows that the internal structure of motifs may change
the critical temperature of the Ising model although the
percolation threshold may be the same (see below).
%%One can see that the critical temperature Eq.~(\ref{Tc-clique}) is larger than
%%$T_c$ given by Eq.~(\ref{asympt-Tc}) for networks with short loops
%%of the same size $\ell$ and the same mean degree $\langle Q \rangle$, see in Fig.~\ref{figTc}.
%%This result shows that the connectivity plays more important role than
%%It is interesting to note that according to Eq.~(\ref{asympt-Tc}), at a given mean degree $Q=2Q_{\ell} \gg 1$  for the Poissonian distribution of hyperdegrees, the critical %%temperature $T_c$ tends asymptotically to the limit which does not depends on the loop size $\ell$ at $\ell \geq 3$.
%%According to Eq. (\ref{Tc-clique}), at a given mean degree $Q=(\ell-1)Q_{\ell}$ the $T_{c}(\ell)$ becomes larger than $T_{c}(\ell=2)$ by $\ell-2$. It %%gives a relative correction of order $O((\ell-2)/Q)$, because $T_c (\ell)/T_c(\ell =2)\approx 1+(\ell-2)/Q $.

Thus in the considered random networks the average branching coefficient $B_{\ell}$ is crucially important. If $B_{\ell} \gg 1$, clustering and finite loops lead to a relatively small shift of the critical temperature in comparison to networks without loops but with the same average branching coefficient. However, in networks with a small branching coefficient of motifs, influence of clustering and finite loops is stronger. Indeed, at $\ell=2$ the critical temperature, Eq.~(\ref{Tc3}), exceeds zero, $T_c >0$, if the average branching coefficient $B_{2}>1$  (for the Poisson distribution this corresponds to the mean degree $\langle Q \rangle >1$). This is due to the fact that only in this case there is a giant connected component [see Eq.~(\ref{pt2})]. In networks with finite loops of size $\ell$, the percolation threshold from Eq.~(\ref{pt2}) is $B_{\ell}=1/(\ell-1)$. Therefore, the phase transition appears at the average branching coefficient $B_{\ell}=1/(\ell-1)$ which is much smaller than 1 if $\ell \gg 1$ (for the Poisson distribution this corresponds to the mean degree $\langle Q \rangle >2/(\ell-1) \ll 1$).
%%It is interesting that in the case of $\ell-$cliques the phase transition appears at average branching coefficient $Q=(\ell-1)Q_{\ell} >1$.

Does clustering influence the critical behavior in the network with motifs? In order to answer this question we consider networks that consist of cliques of a given size $\ell$ with the mean number $\langle Q_{\ell}\rangle$ of $\ell-$cliques attached to a node. It is convenient to assume that the distribution function $P_{\ell}(Q_{\ell})$ of $\ell-$cliques has the following asymptotic behavior: $P_{\ell}(Q_{\ell}) \propto 1/Q_{\ell}^{\gamma}$.
%%It gives exact critical behavior for random uncorrelated tree-like networks.
In Eq.~(\ref{phi}), we use  the
%%following
"effective medium" approximation introduced in Ref. \cite{dgm02}:
\begin{align}
&\sum_{\alpha=1}^{Q_{\ell}-1} h_{M_{\ell}}(\alpha) \approx (Q_{\ell}-1) \langle h_{M_{\ell}} \rangle_{T}
\label{em-approx}
\end{align}
where $\langle h_{M_{\ell}} \rangle_{T}$ is the mean effective field acting on a node from an attached $\ell$-clique. As was shown in Ref.~\cite{dgm02}, this approximation takes into account the most dangerous highly connected nodes and gives exact critical behavior for random uncorrelated tree-like networks (see also Refs.~\cite{lvvz02,Dorogovtsev:dgm08}).
%%As it was shown in \cite{dgm02,lvvz02}, this approach gives exact critical exponents.
As a result, we obtain the distribution function $\Phi_{\ell}(H)$ of the total field $H$ acting on a node from attached $\ell$-cliques as a function of $\langle h_{M_{\ell}} \rangle_{T}$,
\begin{align}
&\Phi_{\ell}(H) \approx \sum_{Q_{\ell}} P_{\ell}(Q_{\ell})\frac{Q_{\ell}}{\langle Q_{\ell}\rangle} \delta \Bigl(H - (Q_{\ell}-1) \langle h_{M_{\ell}} \rangle_{T} \Bigr).
\label{phi-appox}
\end{align}
Substituting Eqs.~(\ref{psi}) and (\ref{phi}) into Eq.~(\ref{mean -h}), we obtain a self consistent equation for the mean effective field $\langle h_{M_{\ell}} \rangle_{T}$,
\begin{equation}
\langle h_{M_{\ell}} \rangle_{T}=G(\langle h_{M_{\ell}} \rangle_{T}),
 \label{em-eq}
\end{equation}
where $G(\langle h_{M_{\ell}} \rangle_{T})$ is the right hand side of Eq.~(\ref{mean -h}). Critical behavior of the model is determined by analytical behavior of the function $G(h)$ at small $h$ \cite{dgm02}. Analysis of analytical properties of the function $G(\langle h_{M_{\ell}} \rangle_{T})$ at zero magnetic field shows that if the forth moment of degree distribution is finite, i.e., $\langle Q_{\ell}^4 \rangle \equiv \sum_{Q_{\ell}} P_{\ell}(Q_{\ell}) Q_{\ell}^4 < \infty$ (scale-free networks with the degree exponent $\gamma > 5$), then $G(h)=Ah+Bh^3+o(h^3)$ where $A$ and $B$ are certain functions of temperature $T$. Solving Eq.~(\ref{em-eq}), we find that the spontaneous magnetization $m(T)$ and the mean effective field $ \langle h_{M_{\ell}} \rangle_{T} $ behave as $m(T)\propto \langle h_{M_{\ell}} \rangle_{T} \propto (T_{c}-T)^\beta$ below $T_c$ with the standard mean-field exponent $\beta=1/2$. If $\langle Q_{\ell}^4 \rangle$ diverges but the second moment $\langle Q_{\ell}^2 \rangle$ is finite (scale-free networks with $3 < \gamma \leq 5$) then $G(h)=Ah+B'h^{\gamma-2}+o(h^{\gamma-2})$. In this case, the critical exponent $\beta$ becomes dependent on the asymptotic behavior of the distribution function $P_{\ell}(Q_{\ell})$, namely, $\beta=1/(\gamma-3)$. Finally, if $\langle Q_{\ell}^2 \rangle$ diverges but the mean degree $\langle Q_{\ell} \rangle$ is finite (scale-free networks with $2 < \gamma \leq 3$), then in the thermodynamic limit the critical temperature $T_c$ tends to infinity, i.e., at any finite temperature the Ising model is in the ordered phase.  This is the critical behavior that was found for the ferromagnetic Ising model on random uncorrelated complex networks \cite{dgm02,lvvz02,Dorogovtsev:dgm08}.

Our approach is explicit for sparse random networks that have local hypertree-like structure.
%%This is the basic property assumed in our approach.
Two motifs can only overlap each other by a single node. If two clusters have common edges, one can combine them and introduce a new
%%complex
motif.
%%Internal structure of motifs is not important for critical behavior of the ferromagnetic Ising model.
The internal structure of motifs does not influence the critical behavior of the ferromagnetic Ising model.
However this structure may be essential for models with frustrations (spin glasses). Relaxing the hypertree-like assumption may change crucially the critical behavior of the Ising model. Furthermore, as we have showed above, asymptotic behavior of distribution function of motifs
%%plays an important role in
is significant for the critical behavior of the Ising model and other models of statistical physics on complex networks \cite{Dorogovtsev:dgm08}. In our analysis of critical behavior we also assumed that there are no correlations in the distribution of motifs over a network. The role of degree-degree correlations in critical behavior for percolation on complex networks was demonstrated in Ref. \cite{gdm08}.
%%Thus, our analysis shows that it is the local hypertree-like structure, asymptotic behavior of distribution function of motifs, and correlations between motifs that are %%important for critical behavior of the Ising model.
Weak clustering in
%%that sense that is defined in Section
the sense of Sec.~II does not influence the critical behavior of the Ising model.
%%However,
The influence of strong clustering demands further investigation.

\section{Percolation threshold}
\label{percolation}

Equation (\ref{Tc0}) permits us to find the percolation threshold in
these loopy networks. Below the percolation threshold, a network
consists of finite clusters and there is no giant component. In this
case, there is no phase transition in the Ising model. This spin
system is in a paramagnetic state at any $T$. At a point of the
birth of a giant component, there is a giant connected cluster and
the critical temperature is $T_{c}=0$. Above the percolation
threshold, the critical temperature is non-zero, $T_{c} > 0$.
Therefore, in a general case for an arbitrary distribution function
$P(Q_{2},Q_{3},Q_{4},...)$, the percolation point is determined by
Eq.~(\ref{Tc0}) at $T=0$. In this case, we have
$F_{\ell}(T=0)=\ell-1$. For example, if there are only loops of size
$\ell$ and $\ell'$ then Eq. (\ref{Tc4}) takes a form,
\begin{equation}
\Bigl[ B_{\ell} - \frac{1}{\ell - 1} \Bigr] \Bigl[ B_{\ell'} -
\frac{1}{\ell'-1} \Bigr]=\frac{\langle
Q_{\ell}Q_{\ell'}\rangle^2}{\langle Q_{\ell}\rangle \langle
Q_{\ell'}\rangle}.
\end{equation}
In the case of $\ell$-loop motifs, Eq.~(\ref{Tc1}) gives the
following criterion for the birth of a giant connected component:
\begin{equation}
(\ell-1)B_{\ell}=1, \label{pt2}
\end{equation}
where $B_{\ell}$ is the average branching. At $\ell=2$, this is the Molloy-Reed criterion for ordinary
uncorrelated random networks \cite{mr95}. The percolation threshold
can be seen in Fig.~\ref{figTc} as a critical value of the mean
degree $\langle Q \rangle$, Eq.~(\ref{m-degree}), below which $T_c$
is zero. Our results about the percolation threshold agree with
results obtained in Refs. \cite{newman:n09,kn10b,hmg2010} by use of
different approaches. It is interesting that we have
$F_{\ell}(T=0)=\ell-1$ for both finite loops and cliques of the same
size $\ell$. Therefore, the percolation threshold
Eq.~(\ref{pt2}), i.e., the point of birth of the giant component, is
also the same.

\section{Application to real networks}
\label{real net}
%%%%%%%%%%%%%%%%%%%%%%%%%%%%%%%%%%%%%%%%%%%%%%%%%%%%%%%%%%%%%%%%%%%%
%%%%%%%%%%%%%%%%%%%%%%%%%%%%%%%%%%%%%%%%%%%%%%%%%%%%%%%%%%%%%%%%%%%%
\begin{figure}[t]
\begin{center}
\scalebox{1.4}{\includegraphics[angle=0]{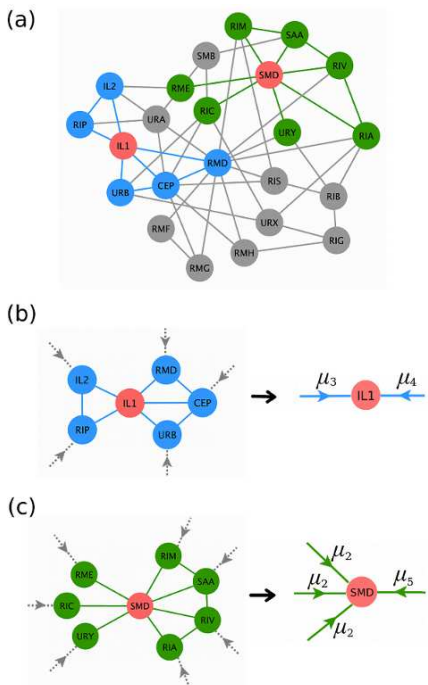}}
\end{center}
\caption{(Color online) %%Motifs in ....An example for applications of belief propagation algorithm to real network.
(a) Small network of motoneurons in \emph{C. elegans} (adapted from Fig.~21(c) in Ref. \cite{white86}). In this figure we follow notations of neurons used in Ref. \cite{white86}. [(b) and (c)] Two subgraphs formed by two nodes IL1 and SMD (red circles) and their nearest neighbors.
(b) The node IL1 has two attached motifs, one triangle and one square.  In (c) the node SMD  has three simple edges and one attached pentagon. Messages sent by these motifs are shown schematically by arrows on the right-hand side.
%%by In order to see the usage of BP algorithm, we choose two nodes IL1 and SMD (Red circles) as an example.
%%The connections between IL1 and the nearest neighbors (SMD) are represented as blue (green) color.
%%(b) Update rule for hyperedges of IL1.
%%Grey dotted arrow means the messages to the nearest neighbor of target nodes through their hyperedges.
%%Two messages $\mu_3$ and $\mu_4$ are transferred to IL1. (b) Update rule for hyperedges of SMD.
} \label{fig5}
\end{figure}
%%%%%%%%%%%%%%%%%%%%%%%%%%%%%%%%%%%%%%%%%%%%%%%%%%%%%%%%%%%%%%%%%%%%
%%%%%%%%%%%%%%%%%%%%%%%%%%%%%%%%%%%%%%%%%%%%%%%%%%%%%%%%%%%%%%%%%%%%

As was mentioned in Sec.~\ref{intro}, real networks are clustered and display network motifs   \cite{ab01a,dm01c,Dorogovtsev:2010,Newman:n03a,Milo2002,Milo2004,Sporn2004,Alon2007}.
%%Finding motifs depends on the kind of network and the functional meaning of motifs. Revealing motifs is non-trivial and very interesting topic of modern investigation.
The ordinary belief-propagation algorithm can give only approximate results for this kind of network. One can improve this approach by use of the generalized belief-propagation algorithm introduced in Sec. \ref{algorithm}. For this purpose,
%%it is necessary to reveal
one should, first, detect motifs attached to each node in a network. As an example we chose a small network of motoneurons in the nervous systems of \textit{Caenorhabditis elegans} \cite{white86}.
%One can apply our method to real network. We explain the application of the BP algorithm to nervous system network in \textit{Caenorhabditis elegans}, (\textit{C. elegans}) %\ref{white}.
The undirected version of this network is shown in Fig.~\ref{fig5}(a). Nodes and links represent neurons and synaptic connections, respectively.
%%The total number of node is $N=23$, the total number of links is $L=51$, and the mean degree of the network is $\left< Q \right> \approx 4.43$.
In order to detect motifs,
%% attached to each node,
we propose the following method. Choose a node and consider a subgraph formed by this node and its nearest neighbors. Only links between nodes in this subgraph are taken into account. The obtained subgraph can be represented as a set of clusters (motifs) overlapping only at the chosen node. Two examples of subgraphs are shown in Figs.~\ref{fig5}(b) and \ref{fig5}(c). One can see that motifs attached to the chosen nodes can be rather complex. If we know motifs attached to every node in the network, then we can find messages sent by these motifs to these nodes using the approach described in Sec.~\ref{algorithm}. Then, using Eq.~(\ref{mean spin 2}), one can find local mean magnetic moments as functions of magnetic field and temperature. In turn, these messages are determined by the update rule, Eq.~(\ref{hyper-ur}).
%%, which uses motifs attached to each node.
Equation (\ref{hyper-ur}) can be solved numerically by using iterations that start from an initial distribution of the messages. This  method allows us to account for clustering in this network [triangles, cliques, and other complex motifs shown, for example, in Figs.~\ref{fig5}(b) and \ref{fig5}(c)]. One can improve this method and find more complex motifs, for example, loops of size 4 or larger, considering subgraphs formed by a chosen node and its first and second nearest neighbors, and so on.

\section{Conclusion}
\label{conclusion}

In the present article, we considered highly structured sparse
networks with arbitrary distributions of motifs and local hypertree-like structure. The considered networks have weak clustering in the sense, that the local clustering coefficient $C(Q)$ decreases faster than the reciprocal degree $1/Q$ (here we use the term \emph{weak clustering} as in Refs.~\cite{Serrano2006a,Serrano2006b,Serrano2006c}). Using the configuration model for
hypergraphs, we introduced a statistical ensemble of these random
networks and found the probability of the realization of the network
with a given sequence of edges, finite loops, and cliques. We
generalized the belief-propagation algorithm to networks with
arbitrary distributions of motifs. Using this algorithm, we solved
the Ising model on networks with arbitrary distributions of
finite loops and cliques. We found an exact critical temperature of
the ferromagnetic Ising model with uniform coupling between spins.
We demonstrated that clustering increases the critical temperature
in comparison with an ordinary tree-like network with the same mean
degree.
%%We showed that networks with loops of large size have a larger critical temperature in comparison with networks having loops of smaller size.
However, weak clustering does not change critical behavior. Considered random complex networks with uncorrelated motifs and weak clustering demonstrate the same critical behavior as random tree-like complex networks. Our solution also enabled us to find the birth
point of the giant connected component in sparse networks with
arbitrary distributions of finite loops and cliques in agreement
with Refs. \cite{newman:n09,kn10b,hmg2010}. We proposed a method how one can account for clustering and find motifs in real networks. We believe that the
proposed generalized belief-propagation algorithm may be used for
studying dynamical processes and variety of models on highly
structured networks with complex motifs.

\begin{acknowledgments}

This work was partially supported by the following PTDC projects:
FIS/71551/2006, FIS/108476/2008, SAU-NEU/103904/2008, and
MAT/114515/2009. S. Y. was supported by FCT under Grant No.
SFRH/BPD/38437/2007.
%%, and also by the SOCIALNETS EU project.

\end{acknowledgments}

%%\newpage

\end{document}